A.M. Chechelnitsky

# GREAT BELT
# OF MEGALITHIC OBSERVATORIES
# AND PROBLEM OF A HISTORY
# OF THE POLE OF THE WORLD

**Chechelnitsky A. M.**

International University Nature, Society and Man "Dubna",
Universitetskaja, 19;
141980, Dubna 3, p.o. box 19, Moscow Region, Russia;
E'mail: ach@thsun1.jinr.ru

**Chechelnitsky A.M.**
**Great Belt of Megalithic Observatories and**
**Problem of a History of the Pole of the World**

### ABSTRACT

The problem of localization of megalithic memorials on the Earth surface is investigated.

It is pointed on existence of *Great Belt of megalithic observatories* - of concentration of astronomically significant objects near geographical latitude $\varphi = 51°$ N.

The latent fundamental (and astronomical) sense is discussed of this phenomenon - the tendency to functional and architectural simplisity, to simmetry of megalithic observatories (in view of the Symmetric Mandale.

It is pointed on possibility of existence of other stretching concentrations - of clusters of megalithic observatories (the Great Belts) in connection with other fixed position of the Poles of the World (of the Earth rotation) in past.

## PROBLEM OF LOCALIZATION
## OF MEGALITHIC STRUCTURES (OBSERVATORIES)

One from the central theme of modern archaeoastronomy (PA, 1974; AA,1993; AC, 1993; AA, 1996; AT, 1996; Atkinson, 1975; etc.) is the next problem:

What for people of the past built such huge monuments as Stonehenge, pyramids in Europe, Asia, America? What purposes The creators of these structures pursued? Whether is casual they are disposed on a surface of the Earth?

The special message for a narrow circle was devoted to this theme - for the actively working researchers - archeologists (professors Shlosser, Ivanishevsky, Uta Berger etc.). It was made by author in time of a conference SEAC 98 ("Astronomy and Culture"), taking place in Dublin of August 31 - September 2, 1998 (Chechelnitsky, 1998a).

Basic ideas were repeated after this in Dubna September 17 on 9 International conferences " Science, Philosophy, Religion" (Theme "Eschatology") in the report Physical Eschatology: a problem "Space - Earth - Man" - as a problem of extremal natural Accidents" (Chechelnitsky, 1998b).

Let's discuss some special aspects of this extensive theme.

## GREAT BELT OF MEGALITHIC OBSERVATORIES

### Two Aspects
Many observable megalithic structure (observatories) are characterized, at least, by two distinguished special circumstances:

∗ *Megalithic observatory - as Symmetric Mandale*.

∗ *Typical (standard) design*.

Known megalithic observatories contain a *rectangular* of astronomically significant directions - cardinal directions, connected with the Sun (equinoxes, solstices) and Moon (limiting points of rise and set of "high" Moon), *inscribed in circle* (of visiring points, megaliths). Such nonaccidental architecture, reminding, we shall speak, Symmetric Mandale, obviously, has also special astronomical sense (Fig.1).

∗ *Distinguished latitude*.

Known megalithic structures (observatories) have the tendency to place in geographical latitude close to

$$\varphi \sim 50 \div 52°N \sim 51°N.$$

Whether it is casual?

In this connection it is interesting to pay attention to the following the bibliographic indication of Wood (Wood, 1981, p. 23) concerning of astronomically significant architecture of Stonehenge (Fig.1):

"...*Piter Newman* has detected, that the *long* sides of *fourangle* are oriented on the *most northern point of set of the "high" Moon*, and if to look in an opposite direction - at *most southern point of it's rise*. He has opened also one more surprising fact: *four basic stone form fourangle*, short and long sides are *perpendicular* one another. To construct *rectangular*, which sides mark *solar and lunar* directions, it is possible only at *Stonehenge latitude*. On other latitudes would be received *parallelogram*."

Discussed aspects of structure and disposition of megalithic observatories from an analytical, astronomical point of view occurs as interconnected and give occasion for the special statement.

**The Suggestion. (Megalithic observatory –
as Symmetric Mandale; A = $\varphi$ Symmetry).**

# The ancient founders of megalithical observatories tried to use most simple - symmetric design - *rectangular* (of cardinal, astronomically significant directions, connected with the Sun and Moon), inscribed in a circle (of visiring points - stones, megalithes).

# This case corresponds to the special condition

$$A = \varphi,$$

where

$\varphi$ - latitude of monument (observatory),

A (and h) - azimuth (and height) of Sun above true horizon at the moment of rise or set.

Believing, for simplicity, that the seen horizon coincides with true (and then h = 0), we receive from the standard formula of spherical Astronomy

$$\cos A \cdot \cos\varphi \cdot \cos h = \sin\delta - \sin\varphi \cdot \sin h$$

the more simple relation

$$\cos A \cdot \cos \varphi = \sin \delta.$$

Per day of summer solstice is valid $\delta = \varepsilon$, where

$\delta$ - declination of the Sun,

$\varepsilon$ - Inclination of equator to ecliptics,

and then for the considered special case A=φ (A=φ - Symmetry) at this time the ratio

$$\cos^2 \varphi = \sin \varepsilon$$

is valid.

# For those epoch of an ancient history, when the corner not hardly differed from modern

$$\varepsilon = 23°26'21''.448 = 23°.43929 \text{ and}$$
$$\cos^2 \varphi = \sin 23°.43929,$$

observable allowable latitude of localization of megalithic observatories not hardly differed from

$$\varphi \, (=A) = 50°.898526.$$

# This zone of the structurally distinguished latitudes represents everywhere found out in archeological researches the *Great Belt of megalithic observatories*. It extends on all Globe in *Eurasia* - from Atlantic up to Pacific ocean, in *America* – from Pacific ocean - up to Atlantic, i.e. actually has *Global* localization.

Some of the brightest representatives this global Great Belt of megalithic observatories in Eurasia:

megalithic observatories of Southern Ireland, Stonehenge (51°11′ N),

Kievica(Czechia),

Kazarovichi(Ukraine),

Babka (51°),

Hodosevichi (53°),

Tushemla (54°),

Ancient city Arkaim (52°39′),

sacral place Savin on the river Tobol in Transural (Kurgan region) (55.4°),

Mountain Ocharovatelnaya in Western Altai and finding on one latitude with it (51° - 52° N)

Semisart (Kara-Bo) on Altai,

Kurgan - temple Arzhan in Tuva,

Parking Malta in near-Bajkal region.

It is interesting to pay attention also that sacral places the Mongols of epoch Chingiz - khan gravitated to consecrated by ancient legends areas (in area Orhon). They also lay at latitudes of Great Belt of megalithic observatories.

In Western hemisphere the brightest representatives of a Great Belt - "Medicine Wheel" (Vogt, 1993) are disposed on border of USA and Canada. (It is known approximately 135 of Medicine Wheel structures. The greatest concentration them is observed in provinces Alberta and Suscachevan in Canada and in state Montana in USA) .

**OTHER GREAT BELTS**
.
**The Preliminary Analysis of Global Statistics**

Obvious availability of megalithic monuments (observatories) also and at latitudes essentially differing from latitude $\varphi = 51°N$, makes the problem of localization megalithic monuments (observatories) not so unequivocal and considerably more interesting.

**Problem of Clusterization of Set of Monuments**

Let's assume during the further purposeful researches will be revealed, that megalithic monuments have tendency to discretness - to group along any others (not appropriate to latitude 51°N) extended zones on a surface of the Earth.

What such grouping, clusterization of megalithic structures (observatories) can mean?

**Possibilities and Prospects of the New Analysis.**
**Other Poles of Rotation of the Earth -**
**Other Great Belts of Megalithic Observatories**

\# Let during long historical time was saved, cultivated, the tradition did not die to build megalithic monuments (observatories), we shall speak, optimally - by the way of Symmetric Mandale (with use of $A = \varphi$ Symmetry).

\# Then during all this time was realized Unequivocal dependence of latitude $\varphi$ (=A) from $\varepsilon$ - inclination of equator to ecliptics, i.e. the ratio was fair

$$\cos^2\varphi = \sin\varepsilon$$

 # If thus there were long enough epoch, when $\varepsilon$ – the inclination of equator to ecliptics had *others (*not equal modern $\varepsilon = 23°.439$) the fixed significances (i.e. existed other Poles of the World (of rotation of the Earth), that, quite probably, could to exist and *other* fixed Great Belts of megalithic observatories, connected with *other* latitudes $\varphi$ (=A) in system coordinates connected to these poles of rotation of the Earth.

 # Simultaneously there is an objective enough tool researches of evolution of a Pole of rotation of the Earth, using experience of Human history (and not just geophysical, astrophysical data).

## THE PERSPECTIVES OF ANALYSIS OF GLOBAL STATISTICS

Attentive purposeful study of global statistics of megalitic monumenus (observatories) can result to fundamental, is possible, unexpected conclusions having extreme significance not only for anthropology (archaeology, histories of civilizations), but also for exact sciences - astrophysics, geophysics, cosmology.

Within the framework of the concept of the Wave Universe and Wave cosmogeonomy [Chechelnitsky, 1980 -1998] we for a long time expect appearance of the new objective data allowing reliably to verify interesting inquests of the theory. Follows to hope, that an extensive material of a human history - still unsufficiently investigated, and main, it is unsufficiently correct interpreted within the framework of exact sciences, - will allow to come nearer to more adequate understanding of evolution of the Nature and Man.

**REFERENCES**

AA (1993). Archaeoastronomy in the 1990s Edited by Clive Ruggles, Group D Publications, (1993).

AC (1993). Astronomies and Cultures, Edited by Clive Ruggles and Nicholas Saunders, Univ. Press of Colorado, (1993).

AT (1996). Astronomical Traditions in Past Cultures, Edited by V.Koleva D. Kolev. Proc. of I Annual General Meeting of SEAC, Smolian, Bulgaria, 31 Aug.-25 Sept. 1993, Sofia, (1996).

AA (1996). Archaeoastronomy: Problems of Formation, Thesises of Int. Conference, Moscow, (1996).

Atkinson R.J.C., Stonehenge, Pelican, London, (1960).

Atkinson R.J.C., Megalithic Astronomy: a Prehistorian's Comments, J. For the History of Astronomy, 6, p.42-52, (1975).

Chechelnitsky A.M., Extremum, Stability, Resonance in Astrodynamics and Cosmonautics, M., Mashinostroyenie, 312 pp. (1980) (Monograph in Russian); (Library of Congress Control Number: 97121007; Name: Chechelnitskii A.M.).

Chechelnitsky A.M., Wave Structure, Quantization, Megaspectroscopy of the Solar System; In the book:Spacecraft Dynamics and Space Research, M., Mashinostroyenie, pp. 56-76, (in Russian) (1986).

Chechelnitsky A.M., Uranus System, Solar System and Wave Astrodynamics; Prognosis of Theory and Voyager-2 Observations, Doklady AN SSSR, v.303, N5 pp.1082-1088, (1988).

Chechelnitsky A.M., Wave Structure of the Solar System, Report to the World Space Congress, Washington, DC, (Aug.22-Sept.5), (1992).

Chechelnitsky A.M., Wave World of Universe and Life: Space - Time and Wave Dynamics of Rhythms, Fields, Structures, Report to the XV Int. Congress of Biomathematics, Paris, September 7-9, 1995; Bio-Math (Bio-Mathematique & Bio-Theorique), Tome XXXIV, N134, pp.12-48, (1996).

Chechelnitsky A.M., On the Way to Great Synthesis of XXI Century: Wave Universe Concept, Solar System, Rhythms

Genesis, Quantization ″In the Large″, pp. 10-27: In book; Proceedings of International Conference ″Systems Analysis on the Threshold of XXI Century: Theory and Practice″, Moscow, 27-29 February 1996, v. 3, Intellect Publishing Hause, Moscow, (1997).

Chechelnitsky A.M., Phenomena of Discretness, Commensurability, Quantization in Wave Universe and Megalitic Astronomy. Preprint, (1998 a).

Chechelnitsky A.M., Physical Eschatology: Cosmos - Earth - Man Problem - As Problem of Extremal Nature Catastrophes, Proceedings of 9 International Conference ″Science, Philosophy, Religion″ (Theme - ″Eschatology″), Dubna, (1998 b).

Wood J.E., Sun, Moon and Standing Stones, Oxford University Press, Russian Translation: Вуд Дж. Солнце, Луна и древние камни, М., Мир, (1981).

PA (1974). The Place of Astronomy in the Ancient World, Philosophical Transactions, R. Soc. London, A.276, (1974).

Vogt D. Medicine Wheel Astronomy. Astronomies and Cultures, Ed. By C. Ruggles and N. Saunders, University Press of Colorado, p.163-201, (1993)

**Chechelnitsky Albert Michailovich -**

is an astrophysicist, cosmologist,

expert in space research, theoretical physics, theory of dynamic systems,

automatic control, optimization of large systems,

econometrics, constructive sociology, anthropology;

COSPAR Associate: Member of International organization - Committee on Space Research (COSPAR) - Member of B, D, E Scientific Commissions.

(COSPAR - most competent international organization, connected with fundamental interdisciplinar investigations of Space).

Author of the (Mega) Wave Universe Concept.